\documentstyle[12pt,epsfig]{article}
\renewcommand{\baselinestretch}{1.75}
\begin{document}
\title{Electron and hole lifetime in monolayer graphene\\ }
\author{Chih-Wei Chiu,* Chiun-Yan Lin, and Rong-Bin Chen\\
\small Department of Physics, National Kaohsiung Normal University ,Kaohsiung 824, Taiwan\\
\small Department of Physics, National Cheng Kung University, Tainan 701, Taiwan\\
\small General Study Center, National Kaohsiung University of Science and Technology,\\
\small  Kaohsiung, 811, Taiwan\\ }
\renewcommand{\baselinestretch}{1}
\maketitle
\renewcommand{\baselinestretch}{1.6}
\begin{abstract}

Excited conduction electrons, conduction holes, and valence holes in monolayer electron-doped graphene
exhibit unusual Coulomb decay rates. The deexcitation processes are studied using the screened exchange energy.
They might utilize the intraband single-particle excitations (SPEs), the interband SPEs, and the plasmon modes,
depending on the quasiparticle states and the Fermi energies. The low-lying valence holes can decay through
the undamped acoustic plasmon, so that they present very fast Coulomb deexcitations, nonmonotonous energy
dependence, and anisotropic behavior. However, the low-energy conduction electrons and holes are similar to
those in a two-dimensional electron gas. The higher-energy conduction states and the deeper-energy valence
ones behave similarly in the available deexcitation channels and have a similar dependence of decay rate on the
wave vector ${\bf k}$.

\vskip 0.6 truecm
\par\noindent Corresponding author\,:\,giorgio@mail.nknu.edu.tw(Chih-Wei Chiu)

\pagebreak
\end{abstract}
\newpage

\section{Introduction}

There are some theoretical
and
[1-12] experimental
[13-29] studies on the decay rates of the quasiparticle states in the layered graphenes, Bernal graphite, and carbon nanotubes. The electron-electron Coulomb scatterings,\cite{PLA357;401,PRL93;157402,PhysicaE34;658,PRB73;235407,Mahan} as well as the electron-phonon scatterings,\cite{PLA357;401,PhysicaE34;658,PRL102;107007,NatComm5;3257,NatPhys3;36,PRL84;5002,PRL102;127401,NanoLett4;517,Mahan} play the critical roles in the decay rates (the mean free paths) of the excited electron/hole states, especially for the former.
By using the self-energy method\cite{PhysicaE35;212} and the Fermi-golden rule, the theoretical predictions of Coulomb decay rates in monolayer graphene clearly show that they purely come from the $T$-created intraband e-h excitations, and the dependence on temperature and wave vector is very strong.\cite{PLA357;401,PhysicaE34;658} The Coulomb decay rate is much faster the electron-phonon scattering rate. A 2D monolayer graphene sharply contrasts with a 2D electron gas\cite{Mahan} or a 1D carbon nanotube\cite{PhysicaE34;658} in electronic excitations and deexcitations. The 2D superlattice model, corresponding to the effective-mass model of monolayer graphene, is utilized to evaluate the Coulomb decay rates and quasiparticle energies in the 2D doped graphene and the 3D graphite intercalation compounds.\cite{PRB55;13961,PRB34;979} It should be noticed that the state energies of quasiparticles can only be evaluated by the self-energies.\cite{PRB53;1109} In addition to the interband excitations, the doping-induced intraband excitations and acustic/optical plasmons are very effective deexcitation channels.\cite{PLA357;401,PhysicaE34;658} For donor-type systems with ${E_F\sim\,1}$ eV, the excited valence bands present the oscillatory energy dependence, in which the largest energy widths even achieve the order of 0.1 eV because of the strong plasmon decay channels. The highly anisotropic behavior is revealed under the calculations of the tight-binding model, such as, the direction-dependent Coulomb decay rates in doped graphene,\cite{PLA357;401,NJP16;125002,PRB97;195302} silicene and germanene at large wave vectors. There exist certain important differences among these three systems, being attributed to the very different hopping integrals.

There are four kinds of experimental methods in measuring the quasiparticle decay rates/lifetimes of carbon-related ${sp^2}$-bonding systems.
The femtosecond pump-probe spectroscopies are very powerful tools in exploring the ultrafast relation of photoexcited electrons.
They cover the time-resolved photoemission spectroscopy,\cite{PRL102;107007,NatComm5;3257,NatPhys3;36,PRL84;5002,JPSJ73;3479,PRL76;483,PRL87;267402} absorption/transmission/reflectivity,
[16-22, 29] and flourence spectroscopies,\cite{PRL95;197401,PRL92;17740} in which their measurements are very useful in understanding the quasiparticle behaviors of the excited conduction electrons. Moreover, the energy distributions of the ARPES measurements could provide the inelastic scattering rates of the excited valence holes. From the experimental measurements on graphite by the first equipment,\cite{PRL76;483} the quasiparticles above the Fermi level are identified to exhibit the  3D metallic behavior, being consistent with the layer electron gas theory.\cite{PRL59;485} The generation, relaxation, and recombination of non-equilibrium electronic carriers are clearly observed by the second tool.\cite{PRB42;2842}
Furthermore, the fourth method is utilized to confirm the pronounced temperature-dependent decay rate for the excited electrons near $E_F$.
As to carbon nanotubes, the measurements of the first method show the carrier relaxation of the excited electrons in metallic single-walled and multiwalled systems.
The measured decay rate is ${\sim\,1}$ meV for the low energy excited states.\cite{PRL84;5002,JPSJ73;3479}
The second\cite{PRL90;057404,PRL92;117402} and third\cite{PRL95;197401,PRL92;17740} methods are  made on bundled and isolated single-walled carbon nanotubes.
The decay rate of the first (second) conduction band is ${\sim\,1-5}$ meV (${\sim\,0.5-3}$ meV, which depends on the nanotube.\cite{PRB73;235407}
The first geometry, or energy gap is deduced to come from the intraband carrier deexcitations. The first conduction band also shows a smaller decay rate about several percentages of the large one.\cite{PRB73;235407}
This is attributed to the interband recombination or the defect trapping. The temperature-dependent photoluminesence spectra are measured for a very small ${(6,4)}$ nanotube between 48 K and 182 K. The non-exponential behavior might be associated with the non-radiative decay of excitations.
The high-resolution ARPES spectra of doped graphene systems (e.g., potassium-adsorbed graphenes) have confirmed the strong and non-monotonous dependence on wave vector/energy, mainly owing to the isotropic Dirac-cone band structure and the prominent Coulomb interactions of free conduction electrons.\cite{PRL102;107007,NatComm5;3257,NatPhys3;36}

\section{Theories}

The temperature- and doping-dependent Coulomb decay rates of monolayer graphene/a single-walled carbon nanotube could also be calculated from the Fermi-golden rule.\cite{PLA357;401} The bare and screened Coulomb potentials are the critical mechanisms in understanding the electronic excitations and deexcitations. One first considers the Coulomb intreraction between the ${|{\bf k}, h\rangle}$ and ${|{\bf p}, h^{\prime\prime}\rangle}$ states. The square of the bare Coulomb interaction, which includes the band-structure effect, is expressed as

\begin{eqnarray}
|V({\bf k},{\bf q},{\bf p},h,h^{\prime},h^{\prime\prime},h^{\prime\prime\prime})|^{2}
=v_{q}^{2}|\langle{\mathbf k+q},h^{\prime}|e^{i{\mathbf q\cdot{\mathbf r}}}|{\mathbf k},h\rangle
\langle{\bf p-q},h^{\prime\prime\prime}|e^{-i\bf{q\cdot\bf{r}}}|{\bf p},h^{\prime\prime}\rangle|^{2},
\end{eqnarray}
where ${|{\bf k}, h^{\prime}\rangle}$ and ${|{\bf p}, h^{\prime\,\prime\,\prime}\rangle}$ are two final states. The effective Coulomb potential, being characterized by the dielectric function, is given by

\begin{eqnarray}
|V^{eff}({\mathbf k},{\mathbf q},{\mathbf p},h,h^{\prime},h^{\prime\prime},h^{\prime\prime\prime};T)|^{2}
=\biggr|\frac{V({\mathbf k},{\mathbf q},{\mathbf p},h,h^{\prime},h^{\prime\prime},h^{\prime\prime\prime})}
{\epsilon(q,\omega_{de}({\mathbf q});T)}\biggr|.
\end{eqnarray}
$\omega_{de}=E_{\bf k}^{h}-E_{\bf k+q}^{h^{\prime}}$
is the deexcitation energy from the initial $|\mathbf{k},h\rangle$ state to the final $|\mathbf{k+q},h^{\prime}\rangle$ state through the intraband or interband
process. Such potential is sensitive to the direction and magnitude of the transferred momentum.

The effective e-e interactions could be utilized to explore the decay rate due to the inelastic Coulomb scatterings. From the Fermi golden rule, the quasiparticle decay rates of the excited ${|{\bf k}, h)}\}$ state at any temperature is (${\hbar\,=1}$)

\begin{eqnarray}
\frac{1}{\tau({\bf k},h;T)}=
2\pi\sum_{{\mathbf p},{\mathbf q},\sigma}\sum_{h^{\prime},h^{\prime\prime},h^{\prime\prime\prime}}
f(E^{h^{\prime\prime}}_{{\mathbf p}})(1-f(E^{h^{\prime}}_{{\mathbf k+q}}))
(1-f(E^{h^{\prime\prime\prime}}_{{\mathbf p-q}}))
|V^{eff}({\mathbf k},{\mathbf p},{\mathbf q};T)|^{2}\cr
\times\delta(E^{h^{\prime}}_{{\mathbf k+q}}+E^{h^{\prime\prime\prime}}_{{\mathbf p-q}}
-E^{h^{\prime\prime}}_{{\mathbf q}}-E^{h}_{{\mathbf k}}).
\end{eqnarray}
By using the two relations

\begin{eqnarray}
\delta(E^{h^{\prime}}_{\mathbf{k+q}}+E^{h^{\prime\prime\prime}}_{\mathbf{p-q}}
-E^{h^{\prime\prime}}_{\mathbf{q}}-E^{h}_{\mathbf{k}})=
\int_{\infty}^{-\infty}\delta(E^{h^{\prime}}_{\mathbf{k+q}}-E^{h}_{\mathbf{k}}+\omega)
\,\delta(E^{h^{\prime\prime\prime}}_{\mathbf{p-q}}-E^{h^{\prime\prime}}_{\mathbf{p}}-\omega)\,d\omega
\end{eqnarray}
and

\begin{eqnarray}
\frac{{\rm Im}[\chi({\mathbf q},\omega;T)]}{\pi[exp(-\omega/k_{B}T)-1]}=
\sum_{{\mathbf p},\sigma,h^{\prime\prime},h^{\prime\prime\prime}}
f(E^{h^{\prime\prime}}_{\bf p})f(1-E^{h^{\prime\prime\prime}}_{\bf p-q})
|\langle{\mathbf p-q},h^{\prime\prime\prime}|e^{-i{{\mathbf q}\cdot{\mathbf r}}}|{\mathbf p},h^{\prime\prime}\rangle|^{2}\cr
\times\delta(E^{h^{\prime\prime\prime}}_{{\mathbf p-q}}-E^{h^{\prime\prime}}_{{\mathbf p}}-\omega).
\end{eqnarray}
Equation (3) is further reduced

\begin{eqnarray}
\frac{1}{\tau({\bf k},h;T)}=2\sum_{h^{\prime}}\int\int\frac{d\phi\,dq}{2\pi^{2}}\,\,\,\,\,\,\,\,\,\,\,\,\,\,\,\,\,\,\,\,\,\,\,\,\,\,\,
\,\,\,\,\,\,\,\,\,\,\,\,\,\,\,\,\,\,\,\,\,\,\,\,\,\,\,\,\,\,\,\,\,\,\,\,\,\,\,\,\,\,\,\,\,\,\,\,\,\,\,\,\,\,\,\,\,\,\,\,\,\,\,\,\,\,\,\,\,\cr
\times \frac{coth[\omega_{de}({\mathbf q})/2k_{B}T]-tanh[\omega_{de}({\mathbf q})-E^{h}_{\bf k})/2k_{B}T]}{exp(-E^{h}_{\bf k})+1}\cr
\times v_{q}|\langle {\mathbf k+q},h^{\prime}|e^{i{{\mathbf q}\cdot {\bf r}}}|{\mathbf k},h\rangle|^{2}Im\biggr[\frac{-1}{\epsilon({\mathbf q},\omega_{de}({\mathbf q});T)} \biggr]\,\,\,\,\,\,\,\,.
\end{eqnarray}
The Coulomb decay rate is determined by the Fermi-Dirac distribution of the final electronic state (the first term) and the energy loss function (the second term).

\section{Results}
The band structure of monolayer graphene is calculated by the tight-binding model, and it is also successful to use in the other systems with different dimensions. [39, 47-52] 
Monolayer graphene exhibits the feature-rich band structure, mainly owing to the hexagonal symmetry and the nanoscaled thickness. The linear Dirac-cone bands are isotropic for the low doping (${E_F\le\,1}$ eV), while they gradually become parabolic at higher/deeper state energies [Fig. 1(a)]. The Fermi energy/free carrier density determines the main features of electronic excitations and thus dominate the Coulomb decay channels. When the Fermi level is situated at the gapless Dirac points, the excited electrons/holes at zero temperature could decay into conduction/valence band states only by utilizing the interband single-particle excitations. The increment of $E_F$ creates the intraband e-h excitations $\&$ plasmon modes, and induces the drastic changes in the interband single-particle excitations. Such $E_F$-induced Coulomb excitations greatly diversify the decay channels. As for the excited conduction electrons above the Fermi level, the final states during the Coulomb deexcitations only lie between the initial states and the Fermi momentum [a red arrow in Fig. 1(b)], according to the Pauli exclusion principle and the conservation of energy and momentum. The available deexcitation channels, the intraband e-h excitations, make the most important contributions to the Coulomb decay rates for the low-lying conduction electrons, corresponding to the orange part in Fig. 2(a). But when the initial state energy is high, the interband single-particle excitations might become the effective deexcitation mechanisms [discussed later in Fig. 2(b)].
Concerning the excited holes in the conduction band, they could be de-excited to the conduction states [a blue arrow in Fig. 1(c)] through the intraband e-h excitations because of the low deexcitation energies and transferred momenta. On the other hand, the valence holes present two kinds of decay processes: ${v\rightarrow\,v}$ and ${v\rightarrow\,c}$ in Figs. 1(d) and 1(e), respectively. Their available decay channels, respectively, cover [intraband $\&$ interband e-h excitations] and [interband single-particle excitations $\&$ acoustic plasmon modes], corresponding to the blue arrow in Fig. 1(d) and red arrow in Fig. 1(e). Specifically, the latter has the large deexcitation energies at small momenta and is thus expected to exhibit the efficient and unusual Coulomb decay rates.

The Coulomb decay rates strongly depend on the quasiparticle states of ${(\bf k\,,h)}$. Concerning the excited conduction electrons near the Fermi momenta, the ${c\rightarrow\,c}$ intraband decay processes are the only deexcitation channels; that is, the intraband single-particle excitations make the main contributions to such processes (the orange part in Fig. 2); therefore, the Coulomb decay rates monotonously grow with ${E^c-E_F}$, as indicated in Figs. 3(a) and 3(b) by the orange curves. Apparently, the zero decay rates (the infinite lifetimes) appear at the Fermi-momentum states because of the step-like Fermion distribution function at zero temperature. Furthermore, those of the neighboring excited states [${|E^{c,v}-E_F|<0.5 E_F}$] are roughly proportional to ${(E^{c,v}-E_F)^2ln|E^{c,v}-E_F|}$, purely according to the numerical fitting. Also, the analytic derivations have been done for graphite intercalation compounds.\cite{PRB34;2} Such an energy dependence is characteristic of a 2D electron gas.\cite{Mahan,PRB26;4421} This is not surprising, since, when ${E^{c,v}\rightarrow\,E_F}$, the deexcitation energy is essential linear in $q$ whether the energy band has a linear quadratic energy dispersion. Furthermore, the low momentum-frequency intraband single-particle are the only deexcitation channels. It is for such reasons that the widths of the doped graphene and electron gas near the Fermi level share a common character. The similar results are revealed in doped silicene and germanene.\cite{PRB97;195302} In addition, for the Fermi-momentum states, the temperature-dependent Coulomb decay rates display the ${T^2ln[T]}$ behavior, as observed in a 2D electron gas.\cite{Nature437;1330}

The Coulomb decay rates of the higher-energy conduction electrons are sensitive to the anisotropic energy bands [Fig. 1(a)].
Along both the KM and K$\Gamma$ directions [Figs. 3(a) and 3(b)], ${[1/\tau\,]_{e,c\rightarrow\,c}}$ grows with the increase of $|E^c-E^F|$.
The energy dependence on the two high-symmetry directions lies in whether the interband single-particle excitations become the effective deexcitation channels.
The higher-energy electronic states have the stronger energy dispersions along K$\Gamma$ [Fig. 1(a)]; therefore, their deexcitation energies at large transferred momenta are consistent with those of the interband e-h excitations. For example, the conduction state of $E^c=3E_F$ along K$\Gamma$ presents a plenty of deexcitation channels, as indicated by the orange curves in Fig. 4(a) at ${\theta_q = 0^{\circ}}$. Similar results are revealed in different momentum directions, e.g., ${\theta_{q}=30^{\circ}}$ in Fig. 4(c). The effective deexcitation channels cover the intraband and interband e-h excitations. The latter accounts responsible for the enhanced Coulomb decay rates in the high-energy conduction states.

The deexcitation behaviors of the excited holes strongly depend on whether they belong to the conduction or valence states. Concerning the conduction holes, the Coulomb decay rates are almost isotropic, as indicated by the approximately identical ${[1/\tau\,]_{h,c}}$ along KM and K$\Gamma$ [green curves in Figs. 3(a) and 3(b)]. Furthermore, the energy dependence is similar to that of the low-lying conduction electrons (2D electron gas). Such results directly reflect the fact that the intraband single-particle excitations are the only available deexcitation channels, e.g., the pink curves related to the states very close to the conduction Dirac point [Figs. 4(a)$-$4(d)]. As a result of the distorted linear dispersion, the finite decay rate at the Dirac point is different from the zero decay derived from the effective-mass model.\cite{PRB34;2} Specifically, the Coulomb decay rate is local minimum at the K point, indicating that the Dirac-point state is the most stable among all the excited conduction holes [green curves in Fig. 3(a) and 3(b)].

On the other hand, the decay rates of the valence holes display the unique ${\bf k}$ dependence. The valence states slightly below the Dirac point have significant decay rates [purple arrow in Fig. 3(b)], approaching to those from above the conduction ones.
They present only the ${v\rightarrow\,c}$ decay process, in which the deexcitation channels mainly come from the interband single-particle excitations and the undamped acoustic plasmon modes, as indicated by the black and brown symbols in Figs. 4(a) and 4(c).
They create the important difference above and below the Dirac point. With the increasing valence-state energy, two decay channels, ${v\rightarrow\,c}$ and ${v\rightarrow\,v}$, contribute to the Coulomb decay rates simultaneously. Concerning the former, the available range of the strong acoustic plasmon grows and then declines quickly for the low-lying  valence holes, leading to an unusual peak structure in ${[1/\tau\,]_{h,v\rightarrow\,c}}$ at small ${E^v}$$^{,}$s [the red curve in Figs. 3(a) and 3(b)]. For example, the $E^v=0.9$ $E_F$ valence state along KM possesses the widest plasmon-decay range, associated with the black and brown curves in Figs. 4(a) and 4(c), so it could show the fast Coulomb decay [blue arrow in Fig. 3(a)]. The plasmon-induced deexcitations almost disappear for the deeper valence states, e.g., their absence under ${E^v<-1.5E_F}$ along the KM direction The interband e-h excitations also make part of contributions to ${[1/\tau\,]_{h,v\rightarrow\,c}}$, and they dominate the Coulomb decay rates of the deeper-energy valence holes, e.g., the red curves along KM and K$\Gamma$ at Ev $<$ -1.5 $E_{F}$ . Specifically, for the ${v\rightarrow\,v}$ process, the excited valence holes [the blue curves in Figs. 3(a) and 3(b)] behave as the excited conduction electrons (the orange curves) in terms of the k dependence and the deexcitation channels. The intraband single-particle excitations are the dominating mechanisms in determining  ${[1/\tau\,]_{h,v\rightarrow\,v}}$ of the low-lying valence states. They are substituted by the intraband and interband e-h excitations for the deeper valence holes along K$\Gamma$. This is responsible  for the anisotropic Coulomb decay rates under the specific K$\Gamma$ and KM directions.

The effective deexcitation channels is worthy of a detailed investigation. Each excited state experiences the inelastic e-e Coulomb scatterings along any directions, as clearly indicated by the summation of ${\bf q}$ in Eq. (1), where the transferred momentum is a function of $q$ (magnitude) and $\theta_q$ (azimuthal angle in the range of 2$\pi$). Through the specific excitation spectra, it might exhibit several dispersion relations (less than six) in the ${\bf q}$-dependent deexcitation energies for a fixed $\theta_q$ . The main reason is that both Coulomb excitations and energy bands possess the hexagonal symmetry; that is, the excitation spectra are identical for $\theta_q$, ${\theta_q\, + \pi\,/3}$, ${\theta_q\, + 2\pi\,/3}$, ${\theta_q\, + \pi}$, ${\theta_q\, + 4\pi\,/3}$, and ${\theta_q\, + 5\pi\,/3}$. For example, the excited valence hole state, with the highest Coulomb decay rate along KM (K$\Gamma$), displays three (four) independent dispersive functions (blue curves) for ${\theta_q\,=0^\circ}$ (${\theta_q\,=30^\circ}$). The other excited states in Figs. 4(a) and 4(c) exhibit the similar behaviors. The total deexcitation regions cover the ${\theta_q}$-dependent dispersion relations; that is, they strongly depend on the direction and magnitude of ${\bf q}$, as expected from the basic scattering pictures.

Figures 5$-$7 clearly illustrate the wave-vector- and Fermi-level-dependent Coulomb scattering rates. The decay rates of the valence holes exhibit non-monotonous energy dependence along any wave-vector directions, since the composite deexcitation channels cover the intraband $\&$ interband e-h excitations, and the damped $\&$ undamped acoustic plasmons. The strongest Coulomb scatterings, which are dominated by the undamped/damped collective excitations, come to exist for the valence states below the Dirac point [Figs. 5(d), 5(d), and 5(d)]. The valence-state decay rate strongly depends on the direction of ${\bf k}$; the anisotropic decay is more obvious for the Fermi states. This is closely related to the strong anisotropy of the deeper valence band [Fig. 1(a)]. As for conduction holes and electrons, the Coulomb scattering rates, as measured from that of the Fermi-momentum states, present different behaviors.
The former possesses nearly isotropic Coulomb decay rate in the phase diagram due to the low-energy isotropic Dirac cones.
However, the latter exhibit monotonous energy dependence and the anisotropic deexcitations appear only for the high doping cases.

It would be relatively easy to observe the oscillatory energy dependence and  the anisotropic behavior under higher Fermi energies. Electronic excitations and Coulomb decay rates are very sensitive to the changes in the free-carrier densities with the variation of $E_{F}$. The $(q, \omega)$-phase diagram of $1/\tau$ is drastically altered, as clearly shown in Figs. 5$-$7. For example, the less-damped acoustic plasmon and the almost isotropic excitations are revealed at a sufficiently low Fermi level, e.g., excitation spectra at ${E_F\leq0.5}$ eV. These are directly reflected in the Coulomb decay rates [Figs. 5 and 6]. For the higher Fermi levels, the available momentum-frequency deexcitation ranges of the strongest acoustic plasmons and the interband single-particle excitations are greatly enhanced, since they could coexist together [Fig. 7]. This leads to the stronger dependence of decay rates on the state energy and direction of ${\bf k}$, such as, a detailed comparison among those in Fig. 5(d) at ${E_F=0.5}$ eV, Fig. 6(d) at ${E_F=0.25}$ eV and Fig. 7(d) at ${E_F=0.75}$ eV.
The $E_F$-induced significant differences are further illustrated by the Coulomb decay rates of the specific states. For example, the largest decay rates come from the interband hole channels; the efficient decay is mainly contributed by the plasmons. Furthermore, the stability of the conduction/valence Dirac-point states is held even under heavy dopings, e.g., ${E_F=0.75}$ eV in Fig. 7.

In addition, the Coulomb decay rates of excited conduction and valence states in monolayer germanene have been thoroughly investigated in Ref.\cite{PRB97;195302}, as done for graphene systems. The screened exchange self-energy is suitable for the inelastic Coulomb scatterings in monolayer germanene and silicene  with the spin-orbital interactions; that is, the similar calculations could be finished under the accurate framework of the theoretical mode. It should be noticed that the spin-orbital couplings result in the superposition of the spin-up and the spin-down components. However, it does not need to deal with the spin-up- and spin-down-dependent Coulomb decay rates separately, because they only make the same contribution because of the . conservation of spin configurations during the Coulomb scatterings. It is sufficient in exploring the wave-vector-, conduction-/valence- and energy-dependent self-energy. The effective deexcitation channels, which are diversified by the composite effects due to the spin-orbital couplings and carrier doping,  are predicted to cover the intraband $\&$ interband e-h excitations, and the second, third $\&$ fourth kinds of plasmon modes On the theoretical side, the RPA self-energy will be greatly modified under an external electric/magnetic field, mainly owing to the  spin-split energy bands and wave functions, This is worthy of a complete investigation on the theoretical models and the electric- field- and magnetic-field-dependent Coulomb decay rates, being never studied in the previous calculations.

\section{Conclusions}

Monolayer germanene, silicene, and graphene, with the hexagonal symmetries, exhibit the $p_z$-orbital-dominated low-lying band structures. The first system possesses the weakest nearest-neighbor hopping integral and the largest spin-orbital couplings, so that the essential properties are relatively easily tuned by the external factors, such as, the carrier doping, electric field, and magnetic field. Apparently, graphene shows the strongest hopping integral ($\sim$2.6 eV) because of the smallest C-C bond length. This high-symmetry system  presents a pair of linearly intersecting valence and conduction band at the gapless Dirac points under the negligible spin-orbital interactions, in which the isotropic Dirac-cone structure is further used to investigate the rich and unique physical properties. However, there are important differences between germanene and graphene in electronic excitations and Coulomb decay rates. Germanene is predicted to display the strongly anisotropic excitation/deexcitation behaviors, the second and third kinds of plasmons (the available decay channels),  the e-h boundaries due to the spin-orbital couplings, and the $F$-induced splitting of excitation spectra. Such features are absent in graphene. The theoretical calculations have been done for the excited conduction and valence electrons in graphene, indicating the isotropic deexcitation behaviors under the low doping and a vanishing Coulomb decay rate at the Dirac point.\cite{PLA357;401} The predicted Coulomb decay rates in monolayer germanene, silicene and graphene could be directly verified from the high-resolution ARPES measurements on the energy widths of quasiparticle state at low temperatures.\cite{PRL102;107007,NatComm5;3257,NatPhys3;36,JPSJ73;3479,PRL84;5002}

Up to now, no theoretical models can deal with the complicated dynamic charge screenings using the fully exact manners. The extra comments on the random-phase approximation are useful in understanding the theoretical progress. The RPA is frequently utilized to explore the Coulomb excitations and deexcitations of condensed-matter systems, especially for the high carrier densities in 3D, 2D and 1D materials.\cite{PLA357;401,PRB73;235407,PRB74;085406,PLA352;446,PRB34;979,PRB62;8508} This method might cause the poor results at low free carrier densities in certain many-particle properties, mainly owing to the insufficient correlation effects. A plenty of approximate models have been proposed to address/modify the electron-electron Coulomb interactions, e.g., the Hubbard\cite{JPRLSL276;238} and Singwi-Sjolander models\cite{PR176;589,PRB6;875} for electronic excitation spectra, and the Ting-Lee-Quinn model for Coulomb decay rates.\cite{PRL34;870} Concerning the time-dependent first-principles numerical calculations, accompanied with the Bethe-Salpeter equation, are further developed to investigate the electronic excitations and Coulomb decay rates in detail.\cite{PRL102;076803,PRL102;127401} Such calculations could account for the experimental measurements on excitation spectra and energy widths only under the large energy $\&$ momentum scales. However, it would be very difficult to provide much information about the critical mechanisms and physical pictures in determining the significant bare polarization functions, energy loss spectra, and Coulomb decay rates. Whether the calculated results within RPA are suitable and reliable at low energy is worthy of a systematic study.

\centerline {\bf Acknowledge}

We would like to acknowledge the financial support from the Ministry of Science and Technology of the Republic of China (Taiwan) under Grant No. MOST 107-2112-M-017-001.

\begin{figure}[p]
\begin{center}\leavevmode
\rotatebox{0}{\includegraphics[width=0.85\linewidth]{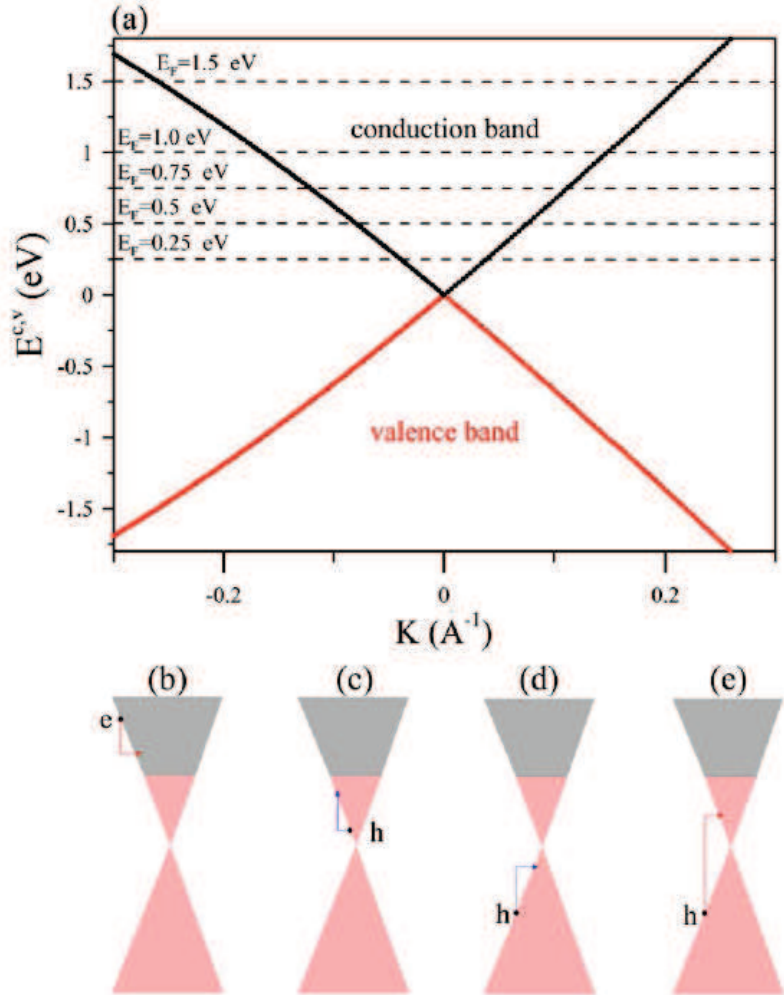}}
\caption{(a) Band structure of monolayer graphene with the various Fermi energies (${E_F=0.25}$ eV, 0.5 eV, 0.75 eV, 1.0 eV and 1.5 eV by the dashed curves)), and the available deexcitation channels
of the specific excited states: (b) conduction electrons , (c) conduction holes, and valence holes with the (d) intraband and (e) interband decays.}
\label{}\end{center}\end{figure}

\begin{figure}[p]
\begin{center}\leavevmode
\rotatebox{0}{\includegraphics[width=0.89\linewidth]{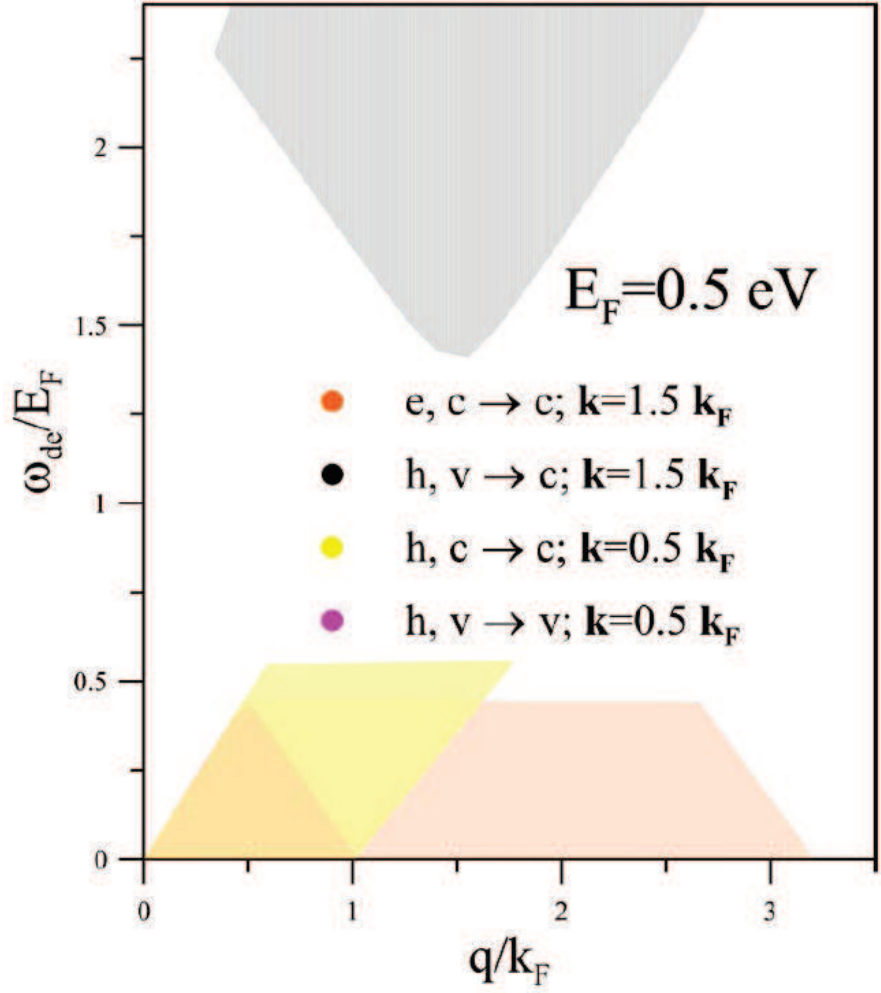}}
\caption{The relations between the deexcitation energies and transferred momenta for the specific excited states; conduction electrons [Fig. 1(b)], conduction holes [Fig. 1(c)], and valence holes [Figs. 1(d) $\&$ 1(e)].}
\label{}\end{center}\end{figure}

\begin{figure}[p]
\begin{center}\leavevmode
\rotatebox{0}{\includegraphics[width=0.88\linewidth]{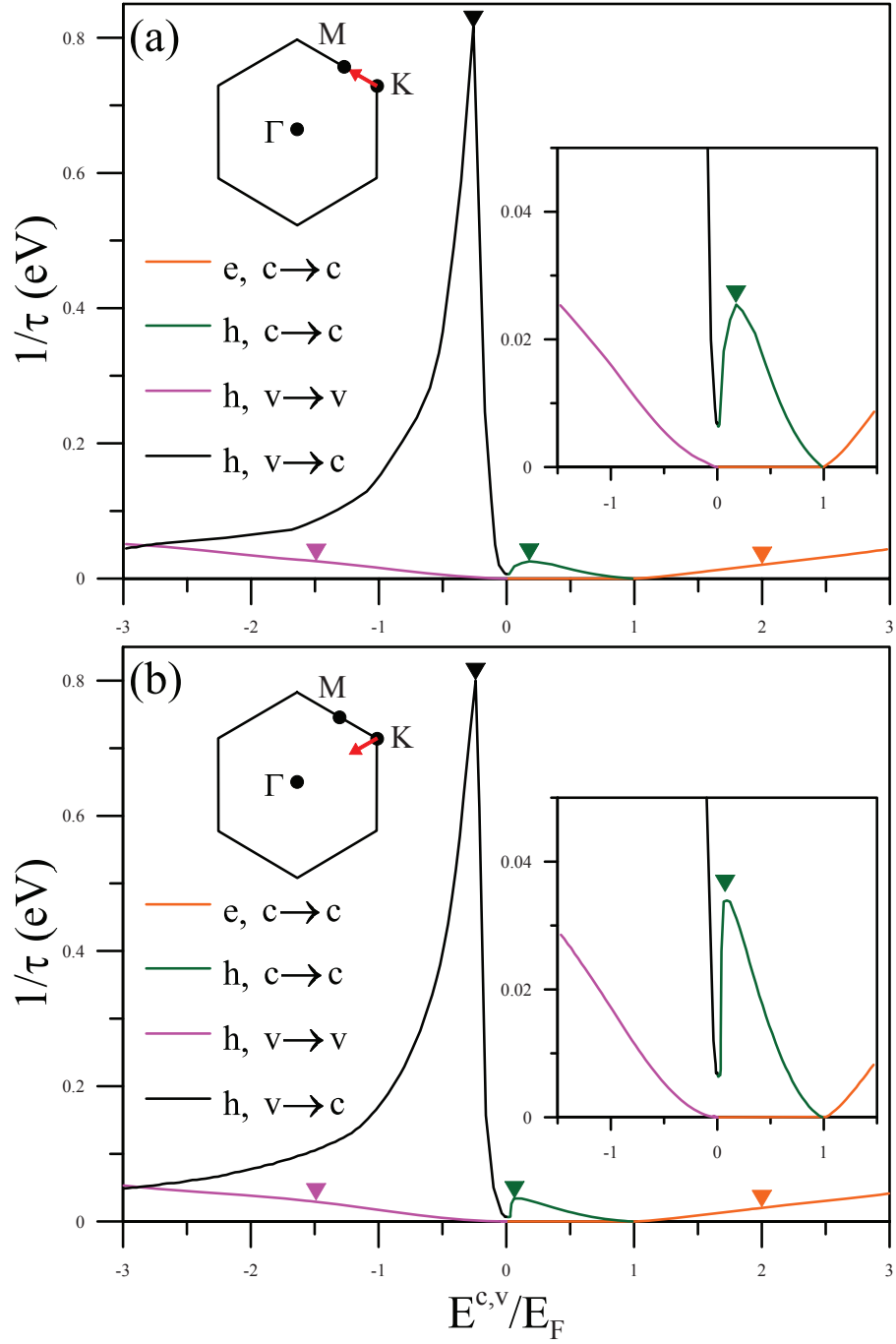}}
\caption{The energy-dependent Coulomb decay rates of the quasiparticle states along the high-symmetry directions: (a) KM and (b) K$\Gamma$ under ${E_F=0.5}$ eV. The insets show the first Brillouin zone.}
\label{}\end{center}\end{figure}

\begin{figure}[p]
\begin{center}\leavevmode
\rotatebox{270}{\includegraphics[width=0.7\linewidth]{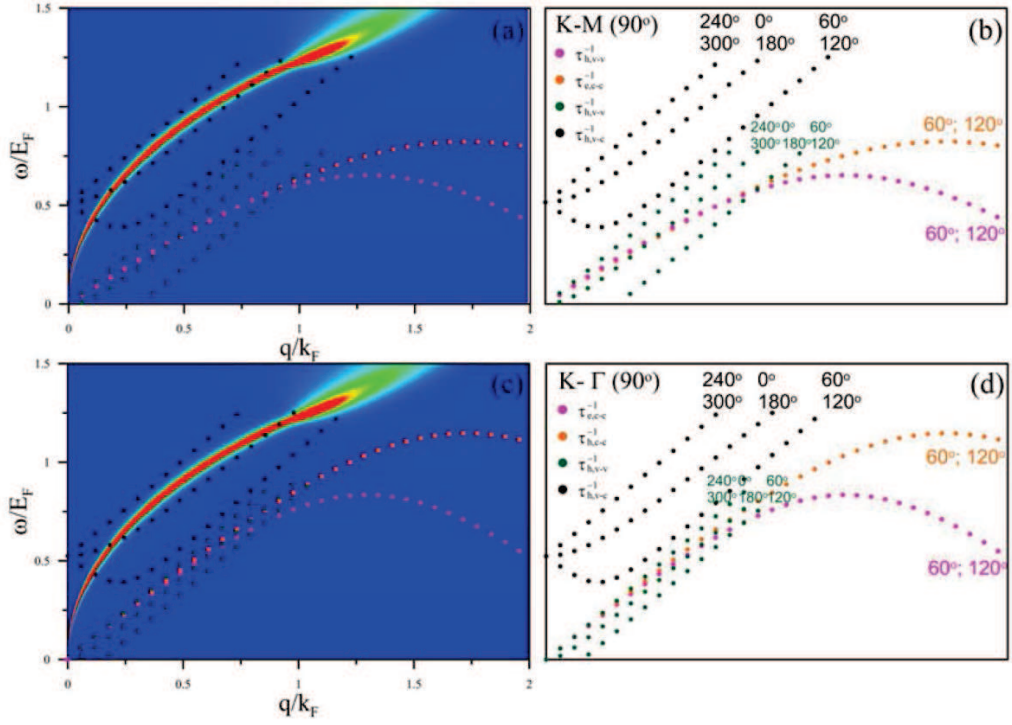}}
\caption{The available deexcitation spectra due to the specific states indicated by the arrows in Figs. 14-6(a) and 14-6(b) are displayed for (a) ${\theta_q\,=0^\circ}$ and (c) ${\theta_q\,=30^\circ}$. (b) and (d) The details of the ${\theta_q}$-dependent deexcitation energies. The curves are defined by the conservation of energy and momentum.}
\label{}\end{center}\end{figure}

\begin{figure}[p]
\begin{center}\leavevmode
\rotatebox{0}{\includegraphics[width=0.9\linewidth]{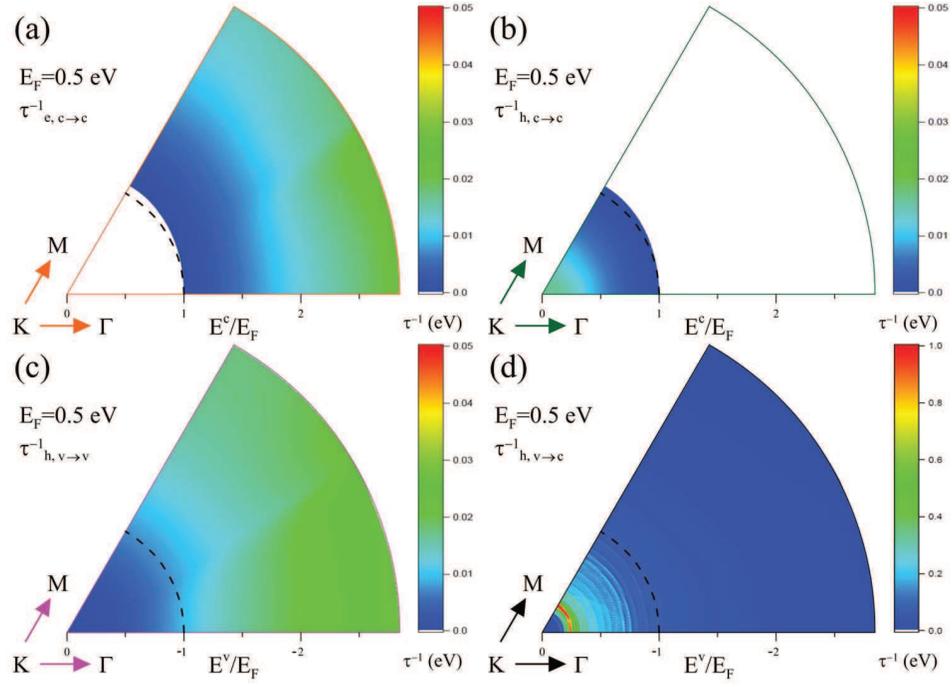}}
\caption{The wave-vector-dependent Coulomb scattering rates of electrons and holes via available deexcitation channels at ${E_F=0.5}$ eV.}
\label{}\end{center}\end{figure}

\begin{figure}[p]
\begin{center}\leavevmode
\rotatebox{0}{\includegraphics[width=0.9\linewidth]{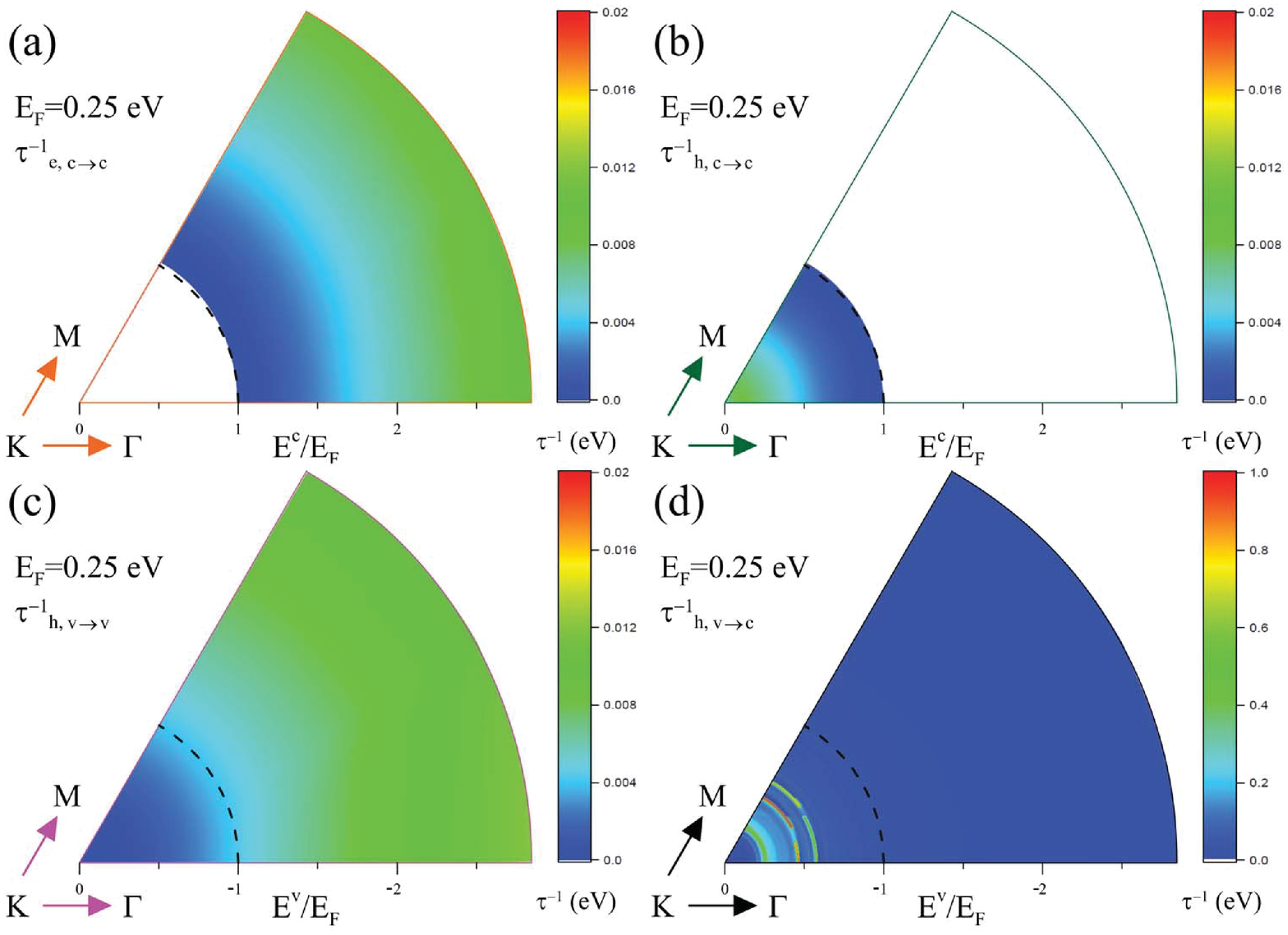}}
\caption{The similar plot as Fig. 5, but shown at  ${E_F=0.25}$ eV.}
\label{}\end{center}\end{figure}

\begin{figure}[p]
\begin{center}\leavevmode
\rotatebox{0}{\includegraphics[width=0.9\linewidth]{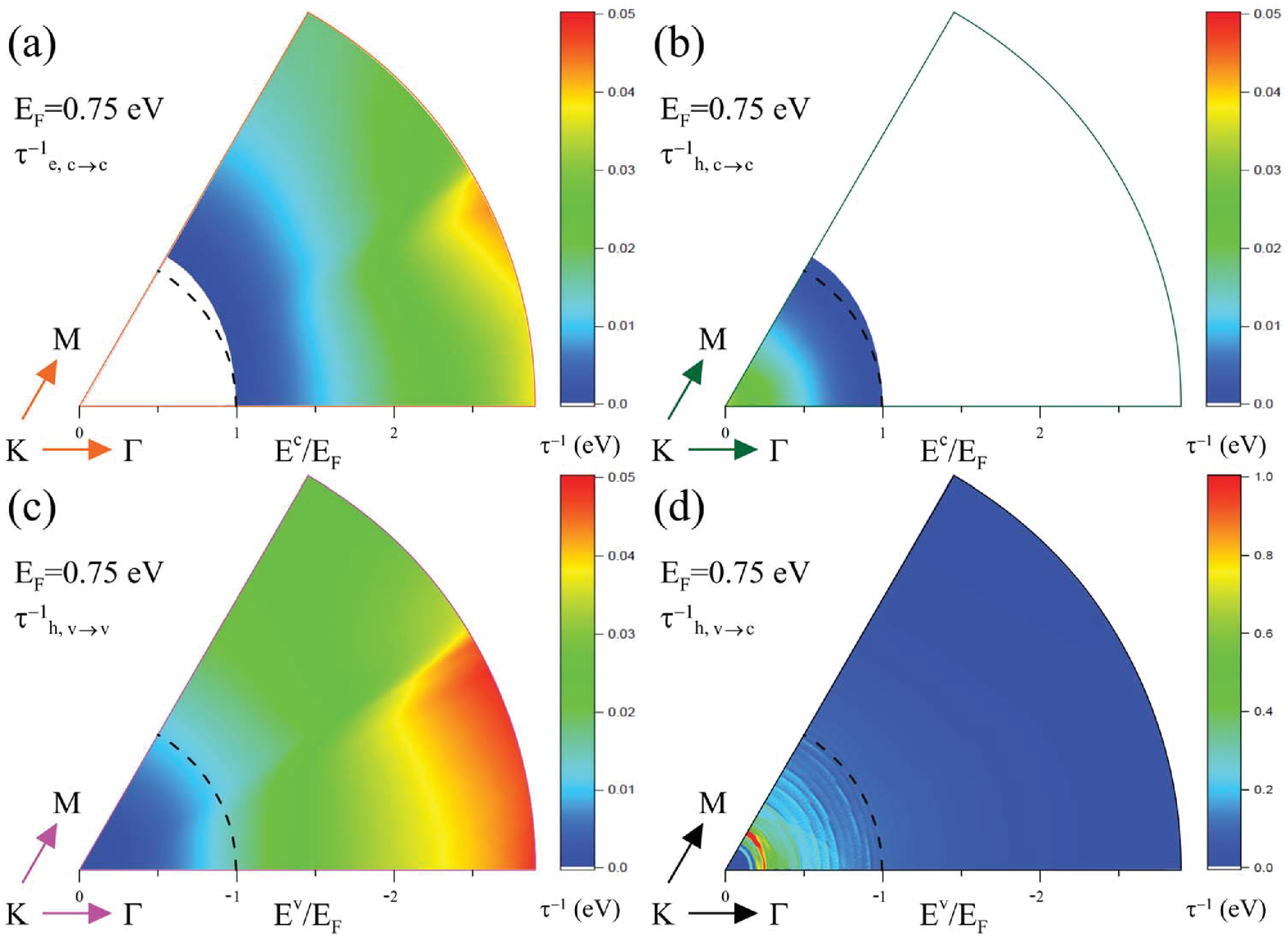}}
\caption{The similar plot as Fig. 5, but shown at  ${E_F=0.75}$ eV.}
\label{}\end{center}\end{figure}

\end{document}